\documentclass[prc,twocolumn,nofootinbib]{revtex4-1}

\usepackage{newtxtext}\usepackage[varvw,bigdelims]{newtxmath}
\usepackage[colorlinks=true,allcolors=blue]{hyperref}
\usepackage[dvipdfmx]{graphicx}
\usepackage{bm}
\usepackage{amsmath}

\begin{document}
% \preprint{YITP-16-6}

\title{Effects of Pauli blocking on pion production in central collisions of neutron-rich nuclei}

\author{Natsumi Ikeno}
\affiliation{Department of Agricultural, Life and Environmental Sciences, Tottori University, Tottori 680-8551, Japan}
\affiliation{Departamento de F\'{\i}sica Te\'orica and IFIC, Centro Mixto Universidad de Valencia-CSIC Institutos de Investigaci\'on de Paterna, Aptdo.~22085, 46071 Valencia, Spain}
\affiliation{RIKEN Nishina Center, 2-1 Hirosawa, Wako, Saitama 351-0198, Japan}
\author{Akira Ono}
\affiliation{Department of Physics, Tohoku University, Sendai 980-8578, Japan}
\affiliation{RIKEN Nishina Center, 2-1 Hirosawa, Wako, Saitama 351-0198, Japan}
\author{Yasushi Nara}
\affiliation{Akita International University, Akita 010-1292, Japan}
\author{Akira Ohnishi}
\affiliation{Yukawa Institute for Theoretical Physics, Kyoto University, Kyoto 606-8502, Japan}

\begin{abstract}
  Pauli blocking is carefully investigated for the processes of $NN \leftrightarrow N \Delta$ and $\Delta \rightarrow N \pi$ in heavy-ion collisions, aiming at a more precise prediction of the $\pi^-/ \pi^+$ ratio which is an important observable to constrain the high-density symmetry energy. We use the AMD+JAM approach, which combines the antisymmetrized molecular dynamics for the time evolution of nucleons and the JAM model to treat processes for $\Delta$ resonances and pions.  As is known in general transport-code simulations, it is difficult to treat Pauli blocking very precisely due to unphysical fluctuations and additional smearing of the phase-space distribution function, when Pauli blocking is treated in the standard method of JAM.  We propose an improved method in AMD+JAM to use the Wigner function precisely calculated in AMD as the blocking probability. Different Pauli blocking methods are compared in heavy-ion collisions of neutron-rich nuclei, ${}^{132}\mathrm{Sn}+{}^{124}\mathrm{Sn}$, at 270 MeV/nucleon. With the more accurate method, we find that Pauli blocking is stronger, in particular for the neutron in the final state in $NN \rightarrow N \Delta$ and $ \Delta \to N\pi$, compared to the case with a proton in the final state.  Consequently, the $\pi^-/\pi^+$ ratio becomes higher when the Pauli blocking is improved, the effect of which is found to be comparable to the sensitivity to the high-density symmetry energy.
\end{abstract}

% \pacs{21.65.Ef, 25.70.-z, 25.80.Ls}
\maketitle
%------------------------------------------------------------
\section{Introduction}

The density dependence of the nuclear symmetry energy is an important information for understanding neutron rich systems such as the nuclear structure, heavy-ion collisions, neutron stars and their mergers~\cite{horowitz2014,abbott2018}. In particular, heavy-ion collisions provide a unique opportunity to study the nuclear equation of state in a wide range of densities, temperatures and neutron-proton asymmetries in the laboratory.

While the neutron-to-proton ratio in the high-density region is not a direct observable in heavy-ion collisions, the $\pi^{-}/\pi^{+}$ ratio of the yields of charged pions can be a sensitive probe of the nuclear symmetry energy at high densities~\cite{bali2008,bali2002,bali2002npa,bali2003}.  In our previous studies~\cite{ikeno2016,ikeno2016erratum}, we calculated the pion production in central collisions of neutron-rich nuclei ${}^{132}\mathrm{Sn}+{}^{124}\mathrm{Sn}$ at 300 MeV/nucleon, using a new approach to combine the antisymmetrized molecular dynamics (AMD)~\cite{ono1992} and a hadronic cascade model (Jet AA Microscopic transport model, JAM)~\cite{nara1999}.  The mechanism of pion production was found to be reflecting the dynamics of neutrons and protons, which are affected by dynamical cluster formation and dissociation (cluster correlation) as well as by the high-density symmetry energy.

At present, some theoretical studies have already been performed with different transport models to investigate the sensitivity of pion observables in heavy-ion collisions~\cite{bali2002,bali2002npa,bali2003,gaitanos2004,reisdorf2007,xiao2009,zqfeng2010,hong2014,guo2014ibuu04}. 
However, some of these results are contradicting to each other even qualitatively, and divergent constraints on the nuclear symmetry energy were obtained so far based on the same experimental data from the FOPI collaboration \cite{reisdorf2007}. 
To solve this kind of problems, the project of the comparison of different transport models was started~\cite{xu2016,yxzhang2018,ono2019} and currently in progress.
Recent works \cite{yxzhang2018,ono2019} in this project performed comparisons under controlled conditions for systems confined in a box, in order to disentangle different sources of uncertainties in the calculated results.

%------Box\\
In Ref.~\cite{yxzhang2018}, different 15 transport codes were compared concentrating on the two-nucleon ($NN$) elastic collision term without mean-field potentials, in a system with an initial Fermi-Dirac distribution at the temperature of either $T = 0$ or 5 MeV.  As an important result, the Pauli blocking factor is found to be affected by the differences among the code strategies to represent the phase space.  The Pauli blocking factor $(1-f)$, or the Pauli blocking probability $f$, is obtained from the phase-space occupation probabilities $f$ in the final state of each scattered nucleon.  Due to the numerical fluctuation associated with the finite number of elements such as test particles, the occupation number $f$ of a phase-space cell can sometimes be larger than 1.  The usual procedure in a case of $f > 1$ is to set $f = 1$, i.e., to completely block the collision.  However, fluctuations to low occupation probabilities are retained.  This decreases the occupation probability on average, i.e.\ $\langle\min(f,1)\rangle\le\langle f\rangle$, leading to overall weaker blocking than in the exact expression. The Pauli blocking effect is therefore underestimated in most of transport codes.  This is a rather fundamental problem in that it is impossible to reconstruct the original distribution $f$ from test particles, which are a finite number of samples taken from $f$.  The problem is more serious in the quantum molecular dynamics (QMD) codes that use one test particle per physical particle to have strong fluctuations than in the BUU codes that use many test particles. However, fluctuations can be of physical importance, because they are used to handle many-body correlations e.g.~to form fragments, in some transport models such as QMD models \cite{aichelin1991} and stochastic mean-field models \cite{colonna1998,napolitani2013}.
%--------------------------

In heavy-ion collisions at several hundred MeV/nucleon, the above-mentioned problem in the Pauli blocking factor may be expected to be not so important because of the high temperature, unlike in the situation of Ref.~\cite{yxzhang2018} for almost degenerate fermionic systems.  However, we recently found that Pauli blocking plays some important role on the pion observables~\cite{ikeno2016erratum}, in the comparison of the results from the AMD+JAM approach in Ref.~\cite{ikeno2016erratum} to those in Ref.~\cite{ikeno2016}.  The Pauli blocking probability $f$ in the $NN\leftrightarrow N\Delta$ and $\Delta\to N\pi$ processes was reduced by a factor 4 in the JAM code employed in Ref.~\cite{ikeno2016}, compared to the proper probability $f$ employed in Ref.~\cite{ikeno2016erratum}.  This affected the results in the studied collisions of neutron-rich nuclei.  The final $\pi^-/\pi^+$ ratio is significantly higher and the numbers of pions and $\Delta$ resonances are smaller in Ref.~\cite{ikeno2016erratum} with the proper blocking probability, than those calculated in Ref.~\cite{ikeno2016} with a reduced blocking probability.
This is understandable because, in neutron-rich systems, neutrons in final states are blocked more strongly than protons and therefore the Pauli blocking for nucleons associated with the $\Delta^{-}$ production is expected to be weaker than that for $\Delta^{++}$ production. The strength of blocking is also important to determine the absolute numbers of produced $\Delta$ resonances and pions in  heavy-ion collisions.  Thus, precise treatment of Pauli blocking is required to reliably predict the pion observables, e.g.\ to constrain the high-density symmetry energy.

In our approach of AMD+JAM \cite{ikeno2016,ikeno2016erratum}, the JAM code is used with one test particle per physical particle.  Therefore, the Pauli blocking in JAM suffers from the above-mentioned problem of too weak blocking due to large fluctuations, as observed in Ref.~\cite{yxzhang2018} in all QMD models, for which any general and fundamental solution is not known. In this paper, we propose a solution in our approach (AMD+JAM), which is possible because we can know a more precise phase-space distribution function in the AMD model.  In previous studies \cite{ikeno2016,ikeno2016erratum}, for the $\Delta$ and pion production processes, we normally used the Pauli blocking factor estimated in the JAM code using test particles.  However, Pauli blocking can be more faithfully treated using the Wigner function calculated from the AMD wave function in the AMD code.  This new method for Pauli blocking and the other methods are formulated and reviewed in Sec.~\ref{sec:formulation}. The results from different Pauli blocking treatments are compared in Sec.~\ref{sec:results} for the pion production in the central collisions of neutron-rich nuclei (${}^{132}\mathrm{Sn}+{}^{124}\mathrm{Sn}$) at the incident energy of 270 MeV/nucleon.  We will see the impacts of Pauli blocking, as well as of the high-density symmetry energy and cluster correlations, on the pion productions.  A summary is given in Sec.~\ref{sec:summary}.

%------------------------------------------------------------
\section{Formulation\label{sec:formulation}}

\subsection{Perturbative treatment of $\Delta$ and $\pi$ production}

In the approach of AMD+JAM, as described in Ref.~\cite{ikeno2016}, a basic assumption is that the $\Delta$ and pion production can be treated as perturbation, e.g.\ in heavy-ion collisions at around 300 MeV/nucleon where the number of $\Delta$ resonances and pions existing at any intermediate time is small compared to the total number of nucleons in the system.  Formally one may multiply a parameter $\lambda$ to the $NN\to N\Delta$ cross section, and then $\Delta$ resonances and pions will appear in the first order of $\lambda$ in the solution of a transport equation.  When we ignore the higher orders of $\lambda$, we assume that the different series of processes started with different $NN\to N\Delta$ processes in the same heavy-ion collision do not influence each other.

In the zeroth order of the perturbation, the system is composed of only nucleons, which is calculated within the AMD model in our approach.  The state at a time of an event is represented by a Slater determinant of Gaussian wave packets
\begin{equation}
\langle\bm{r}|\varphi_j\rangle = e^{-\nu(\bm{r}-\bm{Z}_j/\sqrt{\nu})^2}\chi_{\alpha_j},\qquad j=1,2,\dots,A,
\end{equation}
where the wave packet centroid is denoted by $\bm{Z}_j$ which is a complex vector, and the spin-isospin state takes $\alpha_j\in\{p\uparrow\nobreak,\ p\downarrow\nobreak,\ n\uparrow\nobreak,\ n\downarrow\}$.  The width parameter is chosen to be $\nu=(2.5\ \mathrm{fm})^{-2}$ as usual.  The phase-space distribution, or the Wigner function, corresponding to this Slater determinant is
\begin{multline}
\label{eq:AMD-wigner}
f_{\text{AMD}}^{\alpha}(\bm{r},\bm{p})= 2^3 \sum_{j\in\alpha}\sum_{k\in\alpha}
e^{-2\nu(\bm{r}-\bm{R}_{jk})^2}
e^{-(\bm{p}-\bm{P}_{jk})^2/2\hbar^2\nu}B_{jk}B^{-1}_{kj}
\end{multline}
for each spin-isospin state $\alpha$, with $\bm{R}_{jk}=(\bm{Z}_j^*+\bm{Z}_k)/\sqrt{\nu}$, $\bm{P}_{jk}=2i\hbar\sqrt{\nu}(\bm{Z}_j^*-\bm{Z}_k)$ and $B_{jk}=\langle\varphi_j|\varphi_k\rangle$.  The time evolution of the wave packet centroids $\bm{Z}_j$ is determined by the equations of motion under an assumed effective interaction, and by the stochastic $NN$ elastic collisions where the transition may be allowed to the final states with clusters created.  In the present work, we use the same model and the same set of parameters for the AMD calculation as in our previous work \cite{ikeno2016}.  It should be noted here that the Pauli blocking in $NN$ elastic collisions is always treated within the AMD model.  The time evolution of $f_{\text{AMD}}^\alpha(\bm{r},\bm{p})$ is stochastic, due to $NN$ collisions.

The zeroth order solution $f^{\alpha}_{\text{AMD}}(\bm{r},\bm{p})$ for nucleons can be used to obtain the time evolution of $\Delta$ resonances and pions in a transport model, as
\begin{align}
\frac{\partial f_\Delta}{\partial t}
+\frac{\bm{p}}{\sqrt{m^2+\bm{p}^2}}\cdot\frac{\partial f_\Delta}{\partial\bm{r}}
&=I_\Delta[f_N^{(0)},f_\Delta,f_\pi],
\label{eq:fdelta}
\\
\frac{\partial f_\pi}{\partial t}
+\frac{\bm{p}}{\sqrt{m_\pi^2+\bm{p}^2}}\cdot\frac{\partial f_\pi}{\partial\bm{r}}
&=I_\pi[f_N^{(0)},f_\Delta,f_\pi],
\label{eq:fpi}
\end{align}
which are correct in the first order of the perturbation, with the nucleon distribution $f_N$ in the collision terms replaced by $f^{(0)}_N\equiv f^{\alpha}_{\text{AMD}}(\bm{r},\bm{p})$ of each AMD event.  We ignore potentials for $\Delta$ and $\pi$.  As it is usually done in transport simulations with particles of finite width, we treat the spectral function of $\Delta$ as a distribution of the mass $m$ in $f_\Delta(\bm{r},\bm{p},m)$. The collisions terms $I_\Delta$ and $I_\pi$ in Eqs.~\eqref{eq:fdelta} and \eqref{eq:fpi} include the following contributions from $NN\leftrightarrow N\Delta$ and $\Delta\leftrightarrow N\pi$ processes,
\footnote{
The isospin states of particles are explicitly treated, e.g., $N\in\{p,n\}$, $\Delta\in\{\Delta^{++},\Delta^+,\Delta^0,\Delta^-\}$, and $\pi\in\{\pi^+,\pi^0,\pi^-\}$.  The spin degeneracy factors are $g_N=2$, $g_\Delta=4$ and $g_\pi=1$. The quantity $v'$ is defined by $v'=v^*(E_1^*E_2^*)/(E_1E_2)$, where $E_1^*$ and $E_2^*$ (or $E_1$ and $E_2$) are the energies of the incoming particles in their center-of-mass frame (or in the computational frame), and $v^*$ is the relative velocity of the two particles in their center-of-mass frame.  The decay rate $\Gamma'_{\Delta\to N\pi}$ is that in the computational frame.
}
\begin{align}
I_\Delta^{(N_1N_2\to N_3\Delta)}&=
\frac{g_Ng_N}{(1+\delta_{N_1N_2})g_\Delta}\int\frac{d\bm{p}_3}{(2\pi\hbar)^3}\int d\Omega
\ f_{N_1}f_{N_2}
\nonumber\\
&\quad\times
v'\frac{d\sigma_{N_1N_2\to N_3\Delta}}{d\Omega}
(1-f_{N_3})
,\label{eq:IDelta-NNND}\\
I_\Delta^{(N_1\Delta\to N_3N_4)}&=
-g_N\int\frac{d\bm{p}_1}{(2\pi\hbar)^3}\int d\Omega
\ f_{N_1}f_\Delta
\nonumber\\
&\quad\times
v'\frac{d\sigma_{N_1\Delta\to N_3N_4}}{d\Omega}
(1-f_{N_3})(1-f_{N_4})
,\label{eq:IDelta-NDNN}\\
I_\Delta^{(N\pi\to\Delta)}&=
\frac{g_Ng_\pi}{g_\Delta}\int\frac{d\Omega}{4\pi}f_Nf_\pi v'\sigma_{N\pi\to\Delta}
,\\
I_\Delta^{(\Delta\to N\pi)}&=
-\int\frac{d\Omega}{4\pi}f_\Delta\Gamma'_{\Delta\to N\pi}(1-f_N)
,\label{eq:IDelta-DNpi}\\
I_\pi^{(\Delta\to N\pi)}&=
\frac{g_\Delta}{g_\pi}\int dm_\Delta\int\frac{d\Omega}{4\pi}f_\Delta\Gamma'_{\Delta\to N\pi}(1-f_N)
,\label{eq:Ipi-DNpi}\\
I_\pi^{(N\pi\to\Delta)}&=
-g_N\int dm_\Delta\int\frac{d\Omega}{4\pi}f_Nf_\pi v'\sigma_{N\pi\to\Delta}
.
\end{align}
These terms generally include the Pauli blocking factor $(1-f_N)$ for the nucleon(s) in the final state of these processes.  Similar statistical factors for $\Delta$ and pions are not important because the densities of these particles are very low in the systems studied here. We treat these equations for $\Delta$ and $\pi$ in the JAM code \cite{nara1999}, where the particle distribution functions are represented by point-like test particles, with one test particle for each physical particle in the same way as in QMD models.  The information of nucleons in the AMD code is transferred to the JAM code at every 2 fm/$c$ in the form of test particles $(\bm{r}_1,\bm{p}_1)$, $(\bm{r}_2,\bm{p}_2)$, \dots, $(\bm{r}_A,\bm{p}_A)$ that are generated randomly following $f^\alpha_{\text{AMD}}(\bm{r},\bm{p})$ of Eq.~\eqref{eq:AMD-wigner} as the probability distribution.  Thus, for each realization of a set of test particles, the spin-averaged phase-space distribution in the JAM code is represented by
\begin{equation}
\label{eq:f-JAM}
 f_{\text{JAM}}^{\tau}(\bm{r},\bm{p}) = \frac{1}{2} \times (2\pi\hbar)^3 \sum_{j\in\tau} \delta(\bm{r} -\bm{r}_{j})\delta(\bm{p}-\bm{p}_j),
\end{equation}
for nucleons with the isospin $\tau\in\{p,n\}$.  Within the time span of 2 fm/$c$, the JAM code is run as usual, and collisions and decays will take place according to the order of these events.  After each time span of 2 fm/$c$, the nucleon test particles are replaced by those resampled according to $f^\alpha_{\text{AMD}}(\bm{r},\bm{p})$ at the new time.  Some corrections are considered for the conservations of the baryon number, the charge and the energy \cite{ikeno2016}.

All kinds of quantum effects in AMD from the antisymmetrization are contained in $f^\alpha_{\text{AMD}}(\bm{r},\bm{p})$.  In particular, it is not positive definite, and therefore in the phase-space region of $f^\alpha_{\text{AMD}}(\bm{r},\bm{p})<0$ the probability has to be replaced by zero, which can potentially introduce some inaccuracy of the test-particle representation.  To check the accuracy, as mentioned in Ref.~\cite{ikeno2016}, we compared the density profile for the ground state of the Au nucleus, to find no visible difference between the exact density profile and the ensemble-averaged density calculated from the test-particle representation $f^\tau_{\text{JAM}}(\bm{r},\bm{p})$.  Therefore, we can safely assume that this method of test particles should be sufficiently accurate in highly excited situations during heavy-ion collisions.

\subsection{Methods for Pauli blocking in $NN\leftrightarrow N\Delta$ and $\Delta\to N\pi$\label{sec:pboptions}}

\subsubsection{PB(jam)}

Now for the Pauli blocking for the nucleon(s) in the final state of $NN\to N\Delta$, $N\Delta\to NN$ and $\Delta\to N\pi$ processes, the most natural way within the JAM code is to estimate the blocking factors $(1-f_N)$ in Eqs.~\eqref{eq:IDelta-NNND}, \eqref{eq:IDelta-NDNN}, \eqref{eq:IDelta-DNpi} and \eqref{eq:Ipi-DNpi} by using the information of test particles.  A standard way we employed to obtain the results in Ref.~\cite{ikeno2016erratum} is to use
\begin{equation}
\label{eq:f-jam}
 f_{\text{jam}}^{\tau}(\bm{r},\bm{p}) = \frac{2^3}{2} \sum_{j\in\tau} 
e^{-{(\bm{r} -\bm{r}_{j})^2/2L}}
e^{-{2L(\bm{p}-\bm{p}_j)^2}/ \hbar^2}
\end{equation}
as the blocking probability for a nucleon at the phase-space point $(\bm{r},\bm{p})$ and with the isospin $\tau\in\{p,n\}$.  Particle spins are treated in an averaged way for Pauli blocking.  Since the probability has to be a finite-valued function of $(\bm{r},\bm{p})$, one cannot directly use the phase-space representation $f^\tau_{\text{JAM}}(\bm{r},\bm{p})$ of Eq.~\eqref{eq:f-JAM}.  The function in Eq.~\eqref{eq:f-jam} has been smoothed with a Gaussian function in the phase space with the parameter $L= 2.0\ \text{fm}^2$.  Here and in the following, we will use a suffix ``jam'' (or ``amd'' later) in lower case, to indicate the distribution function used as the blocking probability.  The option to evaluate the Pauli blocking probability with $f^\tau_{\text{jam}}$ is denoted by ``PB(jam)'' in this paper.

The occupation probability $f^\tau_{\text{jam}}$ defined by Eq.~\eqref{eq:f-jam} can be lager than 1.  There are at least two reasons for this.  First of all, the phase-space distribution, which is the Wigner transform of the density matrix in quantum mechanics, is not a quantity that is limited by 1 in general.  The blocking factor in the form of $(1-f)$ may be justified under a local-density approximation, but the factor should be modified in more general situations.  Another more important reason for $f>1$ is, as discussed by Ref.~\cite{yxzhang2018}, the fluctuation due to the sampling of a finite number of test particles.  The original occupation probability in the phase space should be a smooth function, but it cannot be precisely reconstructed from a set of test particles sampled from it, even though some smearing is introduced as in Eq.~\eqref{eq:f-jam}.  In the blocking option PB(jam), we use $\min(f^\tau_{\text{jam}},1)$ as the blocking probability.

\subsubsection{PB(amd)}

The unphysical fluctuation in the blocking factor due to the sampling of test particles can be reduced by using many test particles per physical nucleon, as in BUU codes.  However, in the AMD+JAM approach, an almost equivalent solution is found without using many test particles, because we know in principle the original distribution function of Eq.~\eqref{eq:AMD-wigner} from which test particles were sampled.  Namely, when an $NN\leftrightarrow N\Delta$ or $\Delta\to N\pi$ process is attempted in JAM, a more precise blocking probability, faithful to Eqs.~\eqref{eq:IDelta-NNND}, \eqref{eq:IDelta-NDNN}, \eqref{eq:IDelta-DNpi} and \eqref{eq:Ipi-DNpi} with $f_N=f_N^{(0)}$, can be obtained in AMD as
\begin{multline}
\label{eq:amd-wigner}
f_{\text{amd}}^{\tau}(\bm{r},\bm{p})= \frac{2^3}{2} \sum_{j\in\tau}\sum_{k\in\tau}
e^{-2\nu(\bm{r}-\bm{R}_{jk})^2}
e^{-(\bm{p}-\bm{P}_{jk})^2/2\hbar^2\nu}B_{jk}B^{-1}_{kj}
\end{multline}
for the phase-space point of the nucleon(s) in the final state.  This function $f^\tau_{\text{amd}}$ is the same as Eq.~\eqref{eq:AMD-wigner}, but JAM treats the nucleon spin in an averaged way.  Evidently, $f^\tau_{\text{amd}}$ does not include unphysical fluctuations due to the sampling of test particles, though it includes physical fluctuations originating from stochastic processes in the AMD model.  This option to use $f^\tau_{\text{amd}}$ for the Pauli blocking in the $NN\leftrightarrow N\Delta$ and $\Delta\to N\pi$ processes is denoted by ``PB(amd)'' in this paper.  As mentioned above for similar distribution functions, the value of $f^\tau_{\text{amd}}$ is not limited in the range between 0 and 1.  It is replaced by 1 if $f^\tau_{\text{amd}}>1$ and by 0 if $f^\tau_{\text{amd}}<0$, when it is used as the blocking probability.

In our computation, the $NN\leftrightarrow N\Delta$ and $\Delta\leftrightarrow N\pi$ processes take place always in the JAM code.  However, it communicates bidirectionally with an AMD code that calculates the value of $f^\tau_{\text{amd}}(\bm{r},\bm{p})$ upon every request form the JAM code, using the information on the AMD time evolution stored at every 1 fm/$c$.  The AMD wave function at the time closest to the event time of the $NN\leftrightarrow N\Delta$ or $\Delta\to N\pi$ process is used to evaluate $f^\tau_{\text{amd}}$ for blocking.  Since the rates of $NN\leftrightarrow N\Delta$ and $\Delta\to N\pi$ processes are not so high in the system studied here, the numerical cost for the evaluation of $f^\tau_{\text{amd}}$ is very low.

\subsubsection{PB(amd, jam)}

In the blocking option ``PB(amd, jam)'', the Pauli blocking in an $NN\leftrightarrow N\Delta$ process is treated with $f^\tau_{\text{amd}}$, while that in a $\Delta\to N\pi$ decay is treated with $f^\tau_{\text{jam}}$.  This option may be useful for disentangling the effects of blocking in $\Delta\to N\pi$ from those in $NN\leftrightarrow N\Delta$.

\subsubsection{PB($\frac14$jam)}

It may be useful to first understand the magnitude of the effect of blocking itself, before investigating the differences between the treatments for it.  For this purpose, we will show results obtained when the Pauli blocking is artificially weakened.  The option ``PB($\frac14$jam)'' stands for the case where $\frac14\times f^\tau_{\text{jam}}$ is used as the blocking probability.  The results in the figures of Ref.~\cite{ikeno2016} correspond to this case.  
\subsubsection{PB(amd-h)}

As already mentioned, the blocking probability $f^\tau_{\text{amd}}$ of Eq.~\eqref{eq:amd-wigner} evaluated in AMD is essentially identical to $f^\alpha_{\text{AMD}}$ of Eq.~\eqref{eq:AMD-wigner}, which is the Wigner function for the AMD wave function.  On the other hand, the blocking probability $f^\tau_{\text{jam}}$ of Eq.~\eqref{eq:f-jam} evaluated in JAM does not agree with the test-particle representation $f^\tau_{\text{JAM}}$ of Eq.~\eqref{eq:f-JAM}, because of the smearing in $f^\tau_{\text{jam}}$.  It should also be reminded that the test particles are generated in such a way that the ensemble average of $f^\tau_{\text{JAM}}$ will agree with $f^\alpha_{\text{AMD}}$ and thus $f^\tau_{\text{amd}}$.  Therefore, the distribution of $f^\tau_{\text{jam}}$ on average corresponds to a broader distribution than $f^\tau_{\text{amd}}$.

In order to clarify the effects of this additional smearing of the blocking probability, we will consider an option, called ``PB(amd-h)'', to use the Husimi function corresponding to the AMD wave function,
\begin{align}
f^{\tau}_{\text{amd-h}}(\bm{r},\bm{p})
&=\iint\frac{d\bm{r'}d\bm{p'}}{(\pi\hbar)^3}
e^{-2\nu(\bm{r}-\bm{r'})^2}e^{-(\bm{p}-\bm{p'})^2/2\hbar^2\nu}
f^\tau_{\text{amd}}(\bm{r'},\bm{p'})
\\
\label{eq:amd-husimi}
&=\frac12\sum_{j\in\tau}\sum_{k\in\tau}
e^{-\nu(\bm{r}-\bm{R}_{jk})^2}
e^{-(\bm{p}-\bm{P}_{jk})^2/4\hbar^2\nu}B_{jk}B^{-1}_{kj},
\end{align}
as the blocking probability.  The Husimi function is ensured to have a good property as a probability, i.e.\ $0\le f^\tau_{\text{amd-h}}\le 1$.  However, because of the extra smearing, the distribution is broader than the Wigner function $f^\tau_{\text{amd}}$.

%-------------------------------------------
%-------------------------------------------
\section{Numerical Results\label{sec:results}}
We calculate collisions of ${}^{132}\mathrm{Sn}+{}^{124}\mathrm{Sn}$
at 270 MeV/nucleon for the impact parameters $0<b<1$ fm.  We are going to compare the results with different options for Pauli blocking as explained in Sec.~\ref{sec:pboptions}. For this, we may mainly focus on the results from the calculation of
\begin{enumerate}
\item AMD+JAM with clusters (asy-soft)
\end{enumerate}
in which the cluster formations in the final states of $NN$ collisions in AMD are taken into account, and the SLy4 force \cite{chabanat:1997un} is used as the effective interaction corresponding to a soft density dependence of the symmetry energy ($L=46$ MeV, called `asy-soft').

It is also of our interest to investigate the effects of the density dependence of the symmetry energy and the cluster correlations. Therefore, as we did in Refs.~\cite{ikeno2016,ikeno2016erratum}, we will also show the results of
\begin{enumerate}
\item[2] AMD+JAM with clusters (asy-stiff)
\item[3] AMD+JAM without clusters (asy-soft)
\item[4] AMD+JAM without clusters (asy-stiff)
\end{enumerate}
where the last two cases are calculated without cluster correlations. The effective interaction called `asy-stiff' \cite{ikeno2016} has a stiffer density dependence of symmetry energy ($L=108$ MeV).

\subsection{Phase space distribtuion}\label{subsectA}

% ------------ NN -> ND -------------------
\begin{figure*}
\centering
\includegraphics[width=0.33\textwidth]{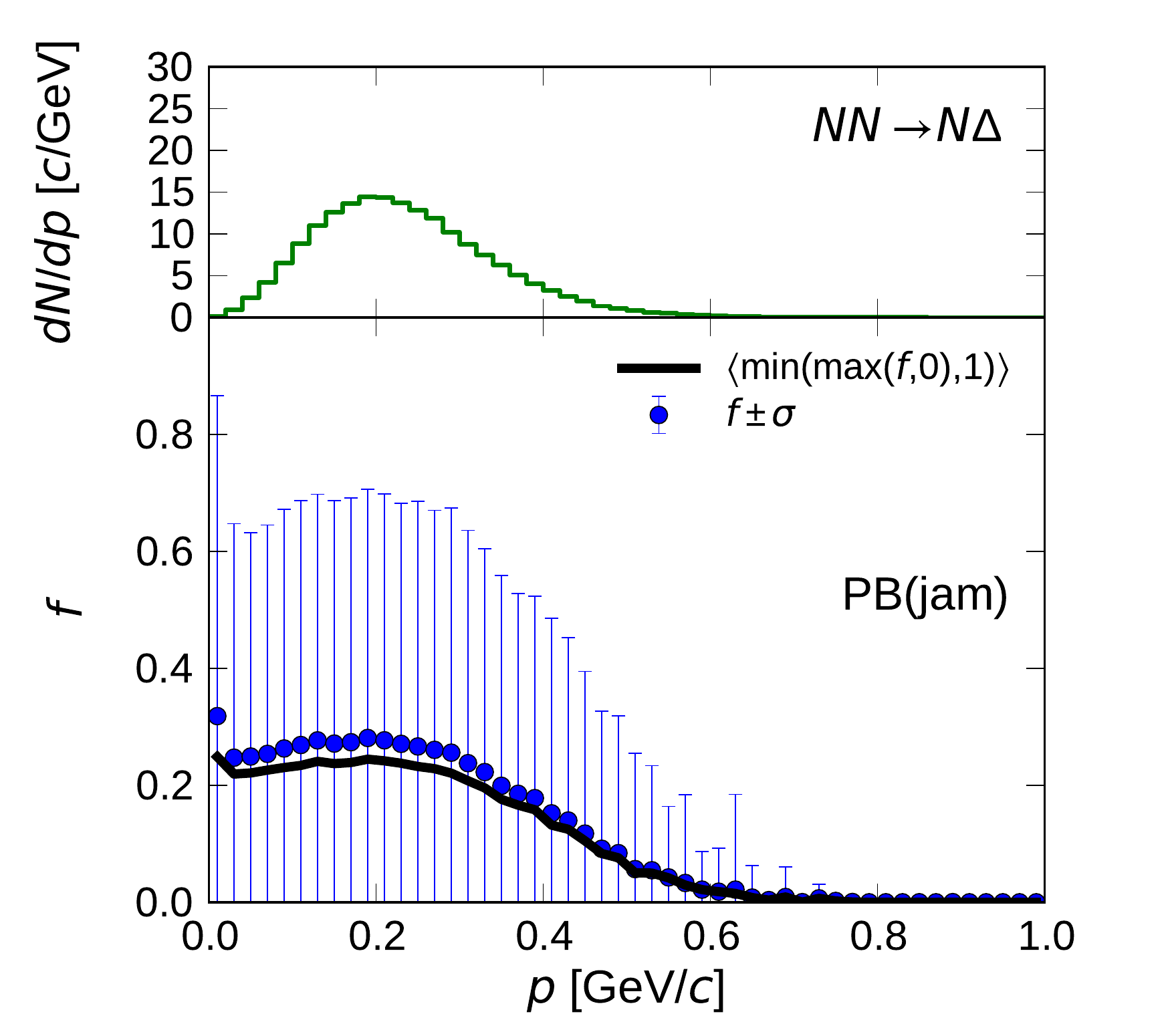}
\includegraphics[width=0.33\textwidth]{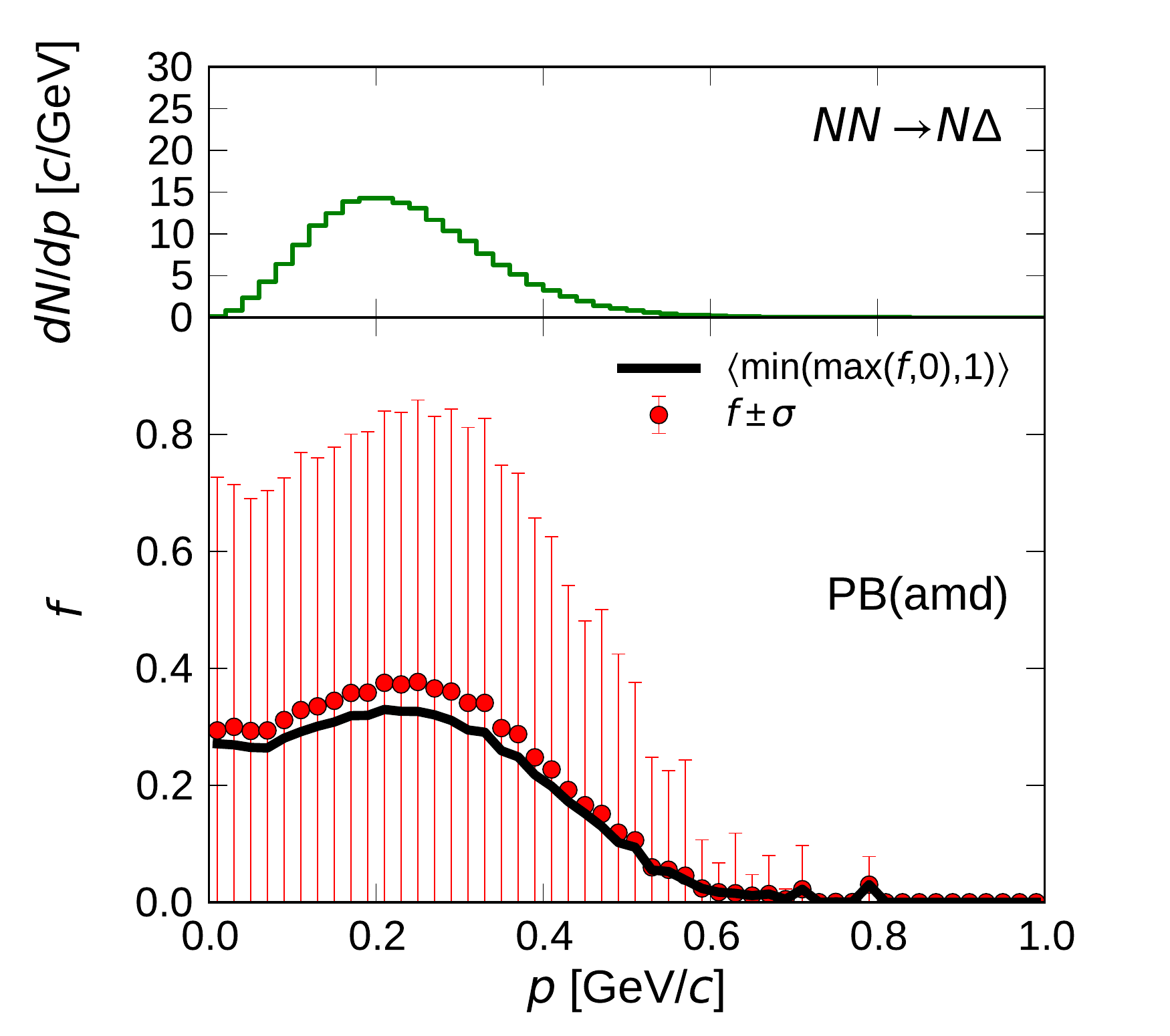}
\includegraphics[width=0.33\textwidth]{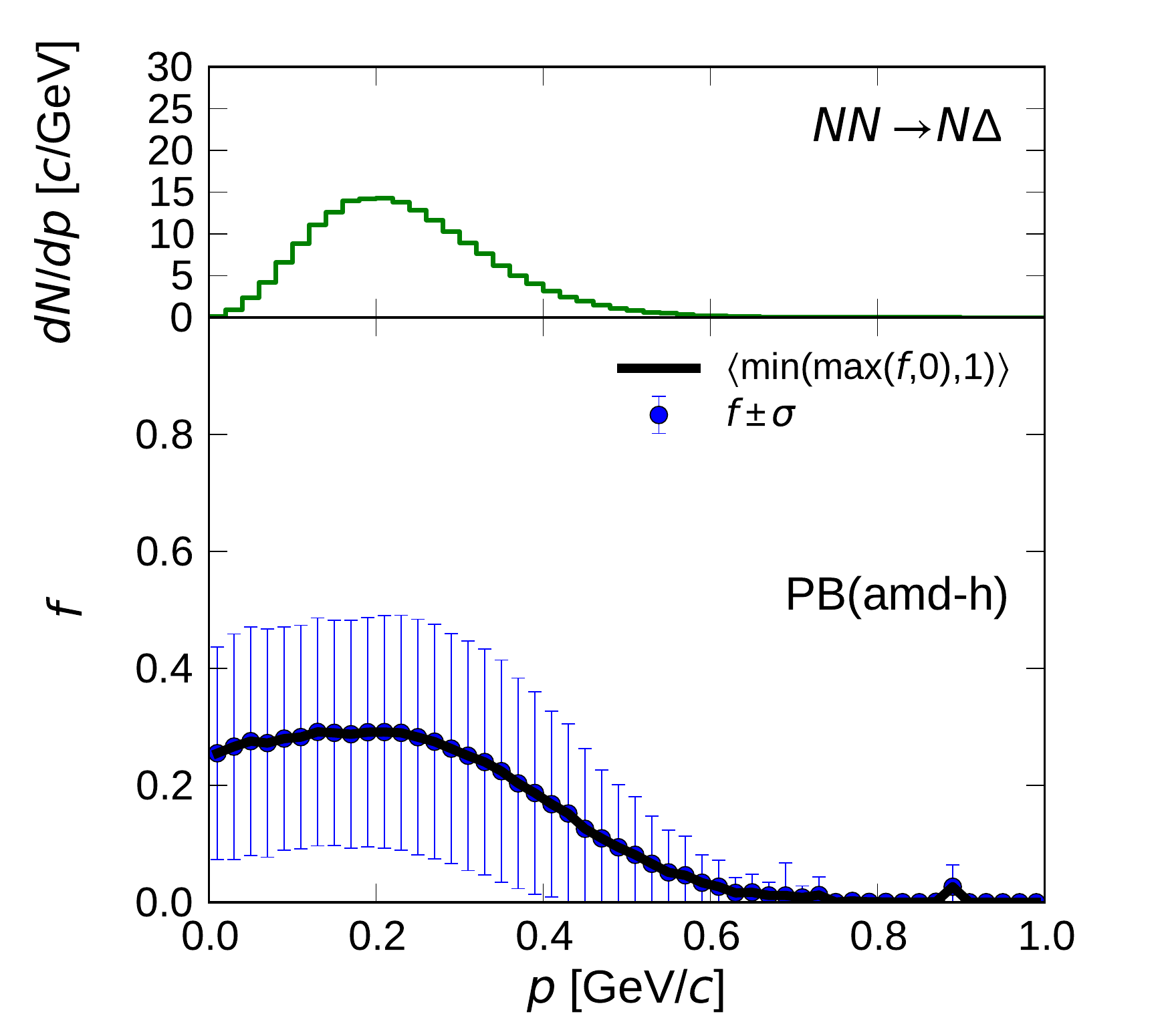}
\caption{\label{fig:f_NNND} Pauli blocking probability for $NN \rightarrow N\Delta$ process in the central collisions of ${}^{132}\mathrm{Sn}+{}^{124}\mathrm{Sn}$ at 270 MeV/nucleon, for the blocking options PB(jam) in the left panel, PB(amd) in the middle panel, and PB(amd-h) in the right panel.  In each panel, the upper part shows the distribution of the momentum $p$ of the final nucleon in the center-of-mass frame of the heavy-ion collision system, and the lower part shows the blocking probability $f=f_{\text{jam}}^\tau$, $f_{\text{amd}}^\tau$ or $f_{\text{amd-h}}^\tau$ as function of the momentum $p$. Points and error bars indicate the mean value $\langle f\rangle$ and its standard deviation.  The actual blocking probability $\langle\min(\max(f,0),1)\rangle$ is shown as the black curve.}
\end{figure*}

First, Fig.~\ref{fig:f_NNND} shows the situation of the Pauli blocking for the final state nucleon in the $NN \rightarrow N \Delta$ process in heavy-ion collisions. The left panel is for the Pauli blocking option PB(jam), which is the standard method of the JAM code to use $f_{\text{jam}}^\tau$ of Eq.~\eqref{eq:f-jam} as the blocking probability. The upper panel of Fig.~\ref{fig:f_NNND}(left) shows the distribution of the momentum $p$ of the final nucleon in the $NN \rightarrow N \Delta$ process, in the center-of-mass frame of the heavy-ion collision. We can see that the momentum $p$ is relatively low, distributed around 0.2 GeV/$c$, because much of the initial $NN$ energy is consumed to change a nucleon to a $\Delta$ resonance. For such a low momentum, we may expect that the Pauli blocking is important, as in fact seen in the lower panel of Fig.~\ref{fig:f_NNND}(left) which shows the blocking probability $f=f_{\text{jam}}^\tau$ as a function of the momentum $p$, in the same way as in Fig.~7 of Ref.~\cite{yxzhang2018}. The mean value $\langle f\rangle$ at a given point of $p$ is shown by a filled blue circle, and the standard deviation of the distribution of $f$ is indicated by the error bar.  As already mentioned above, % Ref.~\cite{yxzhang2018},
the value of probability $f$ sometimes becomes larger than 1.  In this case of PB(jam), we have to truncate $f$ by using $\min(f,1)$ as the blocking probability, and the mean blocking probability $\langle\min(f,1)\rangle$ is shown by the black line in the figure.  Thus, the actual blocking probability is slightly lower than $\langle f\rangle$. It is evident here that the blocking probability is not so large and the difference between $\langle f\rangle$ and $\langle\min(f,1)\rangle$ does not seem so serious as in the case of the degenerate Fermi gas investigated in Ref.~\cite{yxzhang2018}. Nevertheless, we will see later in the results that the observables related to pions are affected by this difference.

The Pauli blocking option PB(amd), shown in the middle panel of Fig.~\ref{fig:f_NNND}, is expected to be the best treatment of our model to use the precise value of the Wigner function calculated in AMD.
In this case, the mean value and the standard deviation of $f=f_{\text{amd}}^\tau$ defined by Eq.~\eqref{eq:amd-wigner} are shown by the point with an error bar for each nucleon momentum $p$. Now it sometimes takes $f>1$ or $f<0$, and therefore $\langle\min(\max(f,0),1)\rangle$ is shown by the black line as the actual blocking probability.
The overall values of $f$ in the option PB(amd) are larger than those in the option PB(jam), especially around $p=0.2$~GeV/$c$.
From this, the Pauli blocking probability in PB(jam) is found to be underestimated, compared to PB(amd) which we believe to be the best treatment of Pauli blocking.

The reduction of $f$ in PB(jam) compared to PB(amd) is most likely due to the extra smearing made in Eq.~\eqref{eq:f-jam} on top of the distribution of the test particles. This interpretation is supported by the right panel of Fig.~\ref{fig:f_NNND} that shows the blocking probability in the option PB(amd-h) which uses Husimi function $f=f_{\text{amd-h}}^\tau$ of Eq.~\eqref{eq:amd-husimi} calculated from the AMD wave function.  As mentioned above, Husimi function is guaranteed that probability $f$ is always between 0 and 1.  For this reason, the mean values $\langle f\rangle$ (blue points) agree with the blocking probabilities (black line).  The fluctuations of $f$ in the option PB(amd-h) are smaller than in PB(amd).  The values of $f$ are lower than those in the option PB(amd) in the region where $f$ is relatively large. These are naturally understood because Husimi function $f_{\text{amd-h}}^\tau$ is obtained by additionally smearing the Wigner function $f_{\text{amd}}^\tau$ in Eq.~(\ref{eq:amd-husimi}).  Also, the overall behavior of $\langle f\rangle$ in PB(amd-h) is very similar to that in the option PB(jam), which is quite reasonable because both $f_{\text{jam}}^\tau$ and $f_{\text{amd-h}}^\tau$ are smeared quantities of $f_{\text{amd}}^\tau$.  The blocking probability $\langle\min(f,1)\rangle$ in PB(jam) is, however, smaller than $\langle f\rangle$ as mentioned above.  The larger fluctuation of $f$ in PB(jam) should be due to the additional fluctuation from the test particle sampling in the AMD+JAM approach.  Thus, as we could expect, Fig.~\ref{fig:f_NNND} shows that the strengths of Pauli blocking in the three blocking options satisfy the relation PB(amd) $>$ PB(amd-h) $>$ PB(jam).

% ------------ ND -> NN, D -> Npi -------------------
\begin{figure}
\centering
\includegraphics[width=0.33\textwidth]{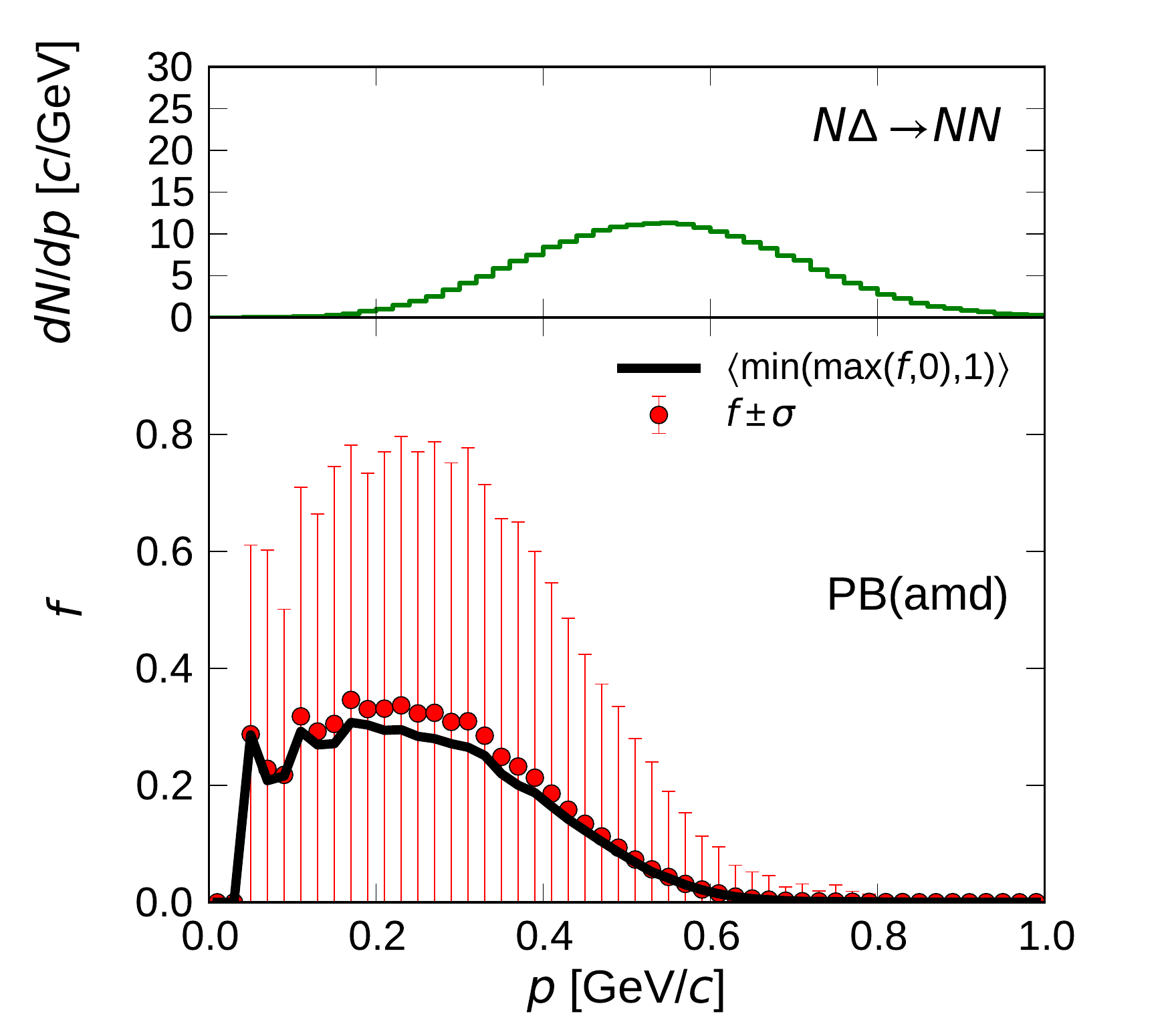}
\caption{ Same as the middle panel of Fig.~\ref{fig:f_NNND}, but for the $N\Delta \rightarrow NN$ process.}
\label{famd_NDNN}
\end{figure}
\begin{figure}
\centering
\includegraphics[width=0.33\textwidth]{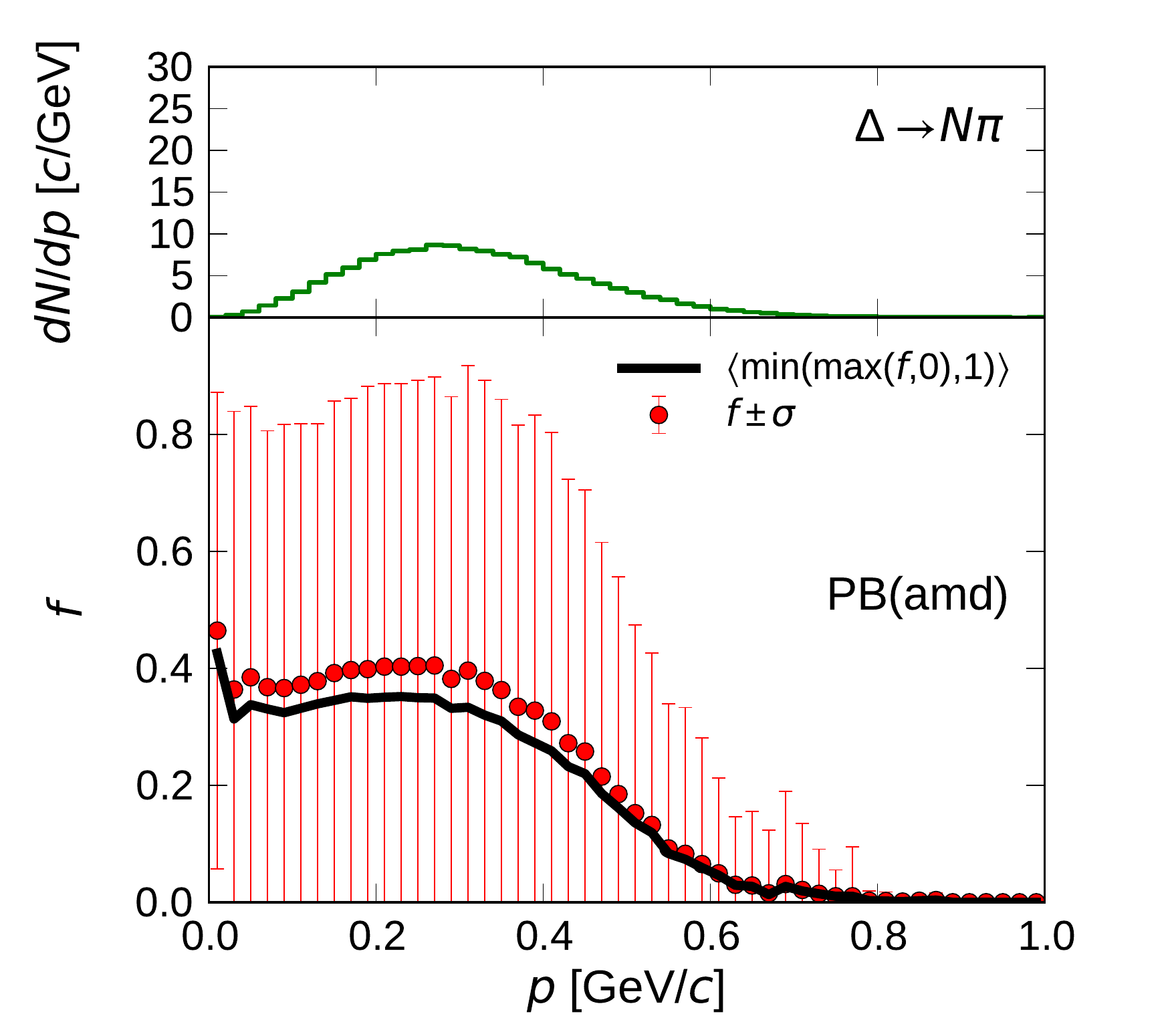}
\caption{ Same as the middle panel of Fig.~\ref{fig:f_NNND}, but for the $\Delta \rightarrow N\pi$ process.}
\label{famd_DNpi}
\end{figure}

Furthermore, in order to see the Pauli-blacking effects in the $N\Delta \rightarrow NN$ and $\Delta \rightarrow N \pi$ processes, we show the results of the option PB(amd) in Figs.~\ref{famd_NDNN} and \ref{famd_DNpi}, respectively.  As seen in the upper panel of Fig.~\ref{famd_NDNN}, the momenta of the final nucleons are relatively high in the $N\Delta \rightarrow NN$ process.  This is because a $\Delta$ resonance has a large mass.  Therefore we can expect that Pauli blocking in the $N\Delta \rightarrow NN$ process is not so important.  On the other hand, in the $\Delta \rightarrow N \pi$ processes of Fig.~\ref{famd_DNpi}, the final nucleon momentum is relatively low, as in the $NN \rightarrow N \Delta$ process.

From these results, the Pauli blocking treatment for the $NN \rightarrow N \Delta$ and $\Delta \rightarrow N \pi$ processes is expected to play an important role in the pion production in heavy-ions collisions. We will discuss the effects in the pion and  $\Delta$ productions in the next subsections.

\subsection{$\Delta$ resonance production}
\begin{figure}
\centering
\includegraphics[width=1.0\columnwidth]{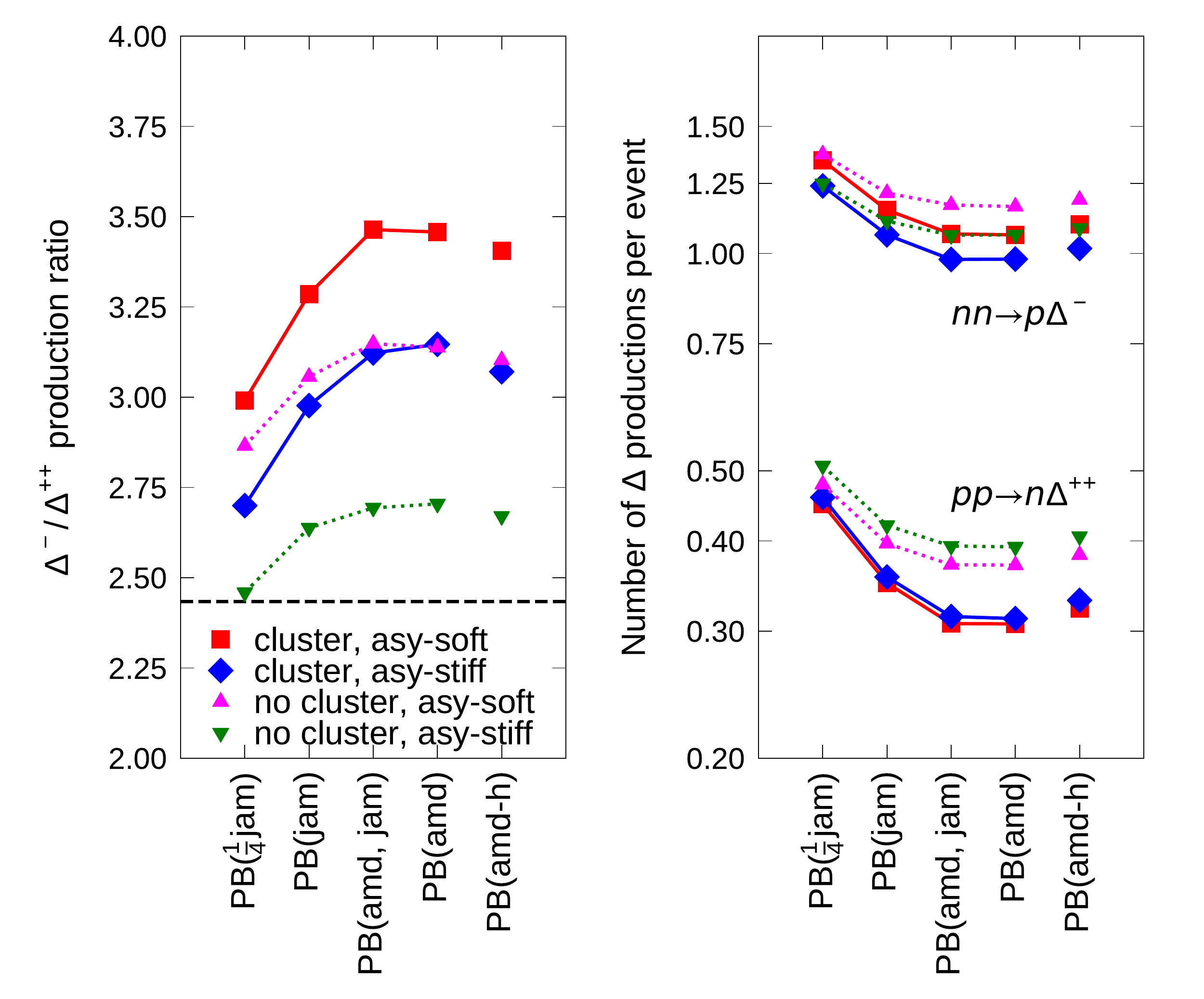}
\caption{\label{fig:deltaprod}
Left panel: $\Delta^{-}/ \Delta^{++}$ ratio of the total production numbers defined in Eq.~(\ref{eq_Delta}) for the different Pauli blocking options. The horizontal dashed line indicates the $(N/Z)^2$ ratio for the total system, $(N/Z)^2_{\text{sys}}=2.4336$.
Right panel: Total production numbers of the $\Delta^-$ and $\Delta^{++}$, for the different Pauli blocking options.  Each symbol corresponds to each of the four AMD calculations for central collisions of ${}^{132}\mathrm{Sn}+{}^{124}\mathrm{Sn}$ at 270 MeV/nucleon. }
\end{figure}

To see the Pauli blocking effect in the $\Delta$ production ($NN \rightarrow N\Delta$) process with different blocking options,
we show in the left panel of Fig.~\ref{fig:deltaprod} a $\Delta^{-}/ \Delta^{++}$ ratio of the total production numbers,
\begin{eqnarray}
 \frac{\Delta^{-}}{\Delta^{++}}=
 \frac{ \int_{0}^{\infty} R(nn \rightarrow p \Delta^{-})dt}{\int_{0}^{\infty} R(pp \rightarrow n \Delta^{++})dt},
\label{eq_Delta}
\end{eqnarray}
where $R(nn \rightarrow p \Delta^{-})$ and $R(pp \rightarrow n \Delta^{++})$ indicate the reaction rates of the $\Delta$ production as functions of time.  The numerator and the denominator of this ratio, namely the total production numbers $ \int_0^\infty R(nn \rightarrow p \Delta^{-}) dt$ and $ \int_0^\infty R(pp \rightarrow n \Delta^{++}) dt$, are shown in the right panel of Fig.~\ref{fig:deltaprod}.  In addition, to see the effects of the different nucleon dynamics in AMD such as due to cluster correlation and the symmetry energy, we show the results from the four different AMD calculations in both panels. Here, we explicitly compare the results of the first three options PB($\frac14$jam), PB(jam) and PB(amd, jam), to see the direct effect of the Pauli blocking for the $\Delta$ production in the $NN \rightarrow N\Delta$ process. In the total production numbers of the $\Delta^{++}$ resonance in the right panel, the effects clearly appear and the numbers are suppressed strongly, as the blocking becomes stronger from PB($\frac14$jam) to PB(jam) and to PB(amd, jam).  It is reasonable that the final neutron in a $pp \rightarrow n \Delta^{++}$ process is blocked strongly in a neutron-rich environment such as in the present system of ${}^{132}\mathrm{Sn}+{}^{124}\mathrm{Sn}$ collisions.  On the other hand, the proton in a $nn\to p\Delta^-$ process may not be blocked so strongly.  Thus, Pauli blocking effect is stronger for the production of $\Delta^{++}$ than for $\Delta^{-}$, and therefore the $\Delta^- /\Delta^{++}$ production ratio in the left panel becomes larger in the stronger Pauli blocking option.
% As mentioned above, the strength of  Pauli blocking option PB(amd) is stronger than that of the PB(jam).
% So, the numbers of the Delta are suppressed (become smaller) in Fig.~\ref{delta_pm}.
We also note that this feature appears strongly in particular when cluster correlation is switched on, which may be expected due to spatial correlations among nucleons in different spin and isospin states.

\subsection{pion production}
\begin{figure}
\centering
\includegraphics[width=1.0\columnwidth]{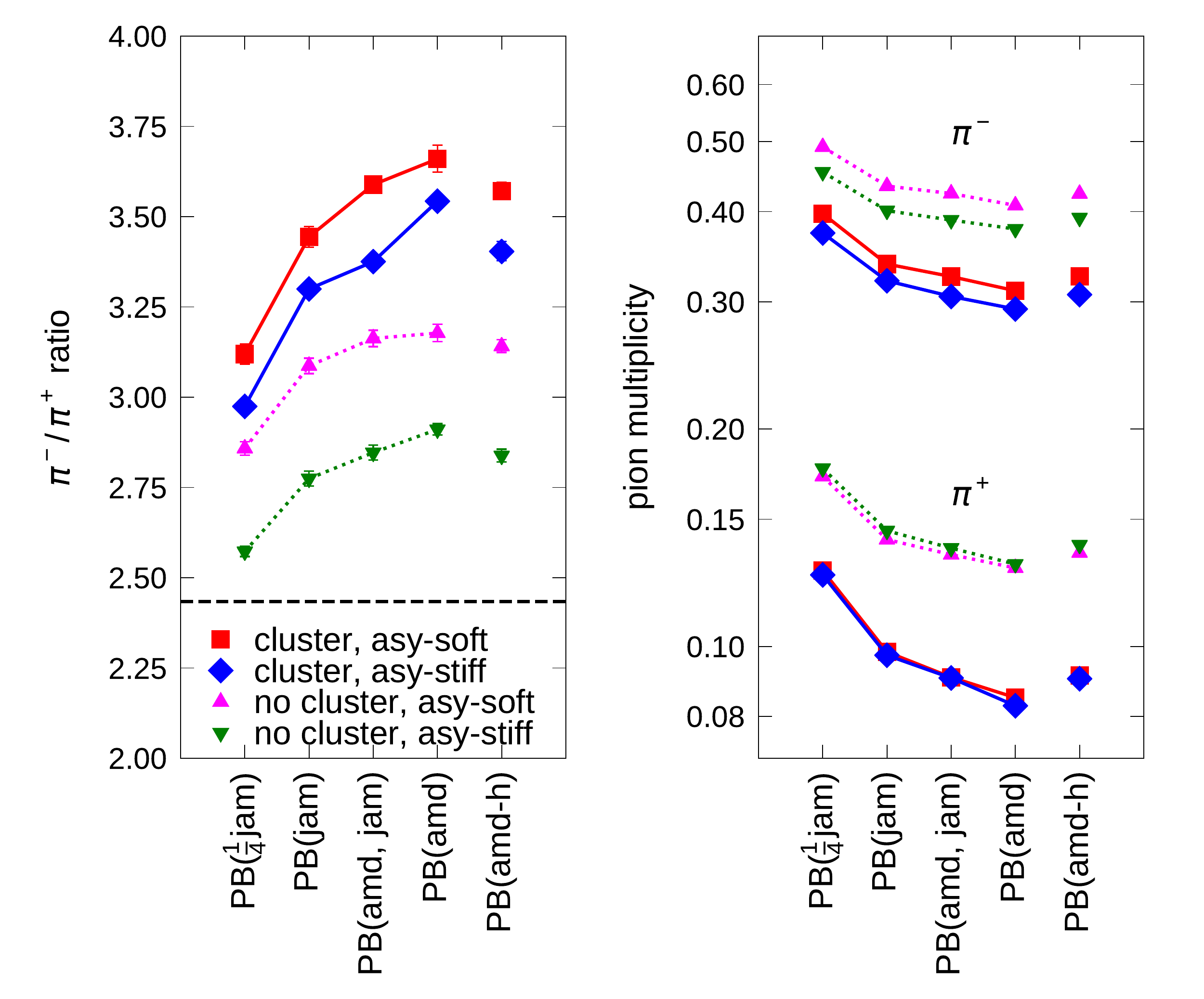}%
\caption{\label{fig:pinum} Left panel: Final $\pi^{-}/ \pi^{+}$ ratio for the different Pauli blocking options. The horizontal dashed line indicates the $(N/Z)^2$ ratio for the total system, $(N/Z)^2_{\text{sys}}=2.4336$.  Right panel: Numbers of $\pi^{-}$ and $\pi^{+}$ in the final state for the different Pauli blocking options.  Each symbol corresponds to the ratio for each of the four cases of calculation for central collisions of ${}^{132}\mathrm{Sn}+{}^{124}\mathrm{Sn}$ at 270 MeV/nucleon. }
\end{figure}

\begin{figure}
\centering
\includegraphics[width=1.0\columnwidth]{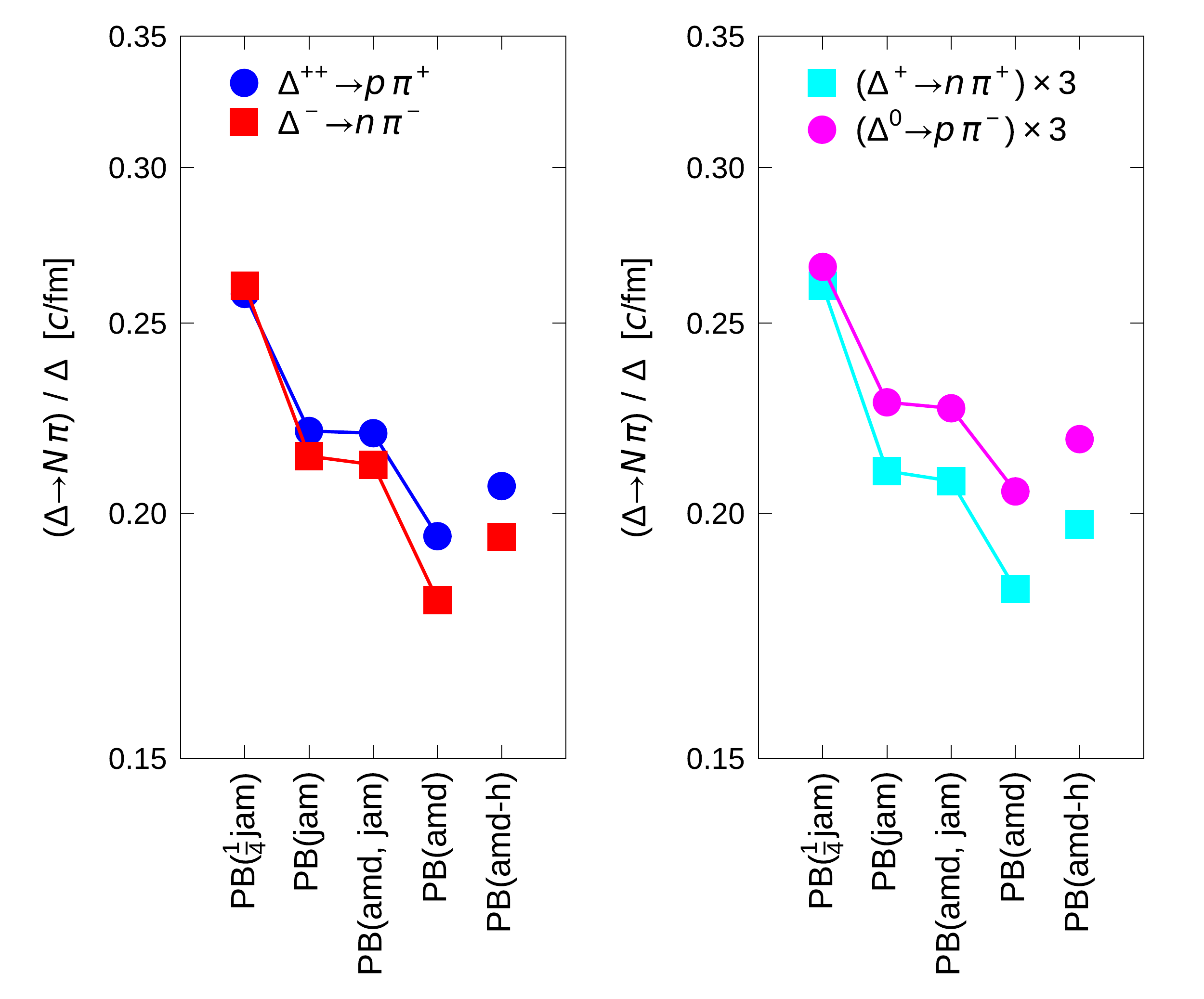}%
\caption{\label{fig:deltadecay}
The integrated $\Delta$ decay rates normalized by the cumulative number of $\Delta$, as defined by Eqs.~\eqref{Delta_pp}, \eqref{Delta_m}, \eqref{Delta_p} and \eqref{Delta_0}, for the different Pauli blocking options.  The left panel shows the quantities for the channels $\Delta^{++}\to p\pi^+$ and $\Delta^-\to n\pi^-$, and the right panel shows the quantities multiplied by a factor 3 for the channels $\Delta^0\to p\pi^-$ and $\Delta^+\to n\pi^+$.
}
\end{figure}

In the left panel of Fig.~\ref{fig:pinum}, we show the pion ratio $\pi^- / \pi^+$ in the final state, for the different Pauli blocking options.
These behaviors of $\pi^-/\pi^+$ in the first three options, i.e.\ PB($\tfrac14$jam), PB(jam) and PB(amd, jam), are quite similar to those of the $\Delta^-/\Delta^{++}$ production ratio found in Fig.~\ref{fig:deltaprod}(left). In fact, this is consistent with our finding in Refs.~\cite{ikeno2016,ikeno2016erratum} that the final $\pi^-/\pi^+$ ratio is strongly correlated with the $\Delta^-/\Delta^{++}$ production ratio in the early stage of the reaction.  The effects in the late stage are that the $\pi^-/\pi^+$ is enhanced when the cluster correlations are taken into account in the AMD calculation, and that the sensitivity to the symmetry energy is somewhat weakened but about 70\% of the sensitivity remains in the final $\pi^-/\pi^+$ ratio.
 Now the main point to be discussed in this subsection is that the $\pi^-/\pi^+$ ratio in Fig.~\ref{fig:pinum}(left) further increases when the Pauli blocking in the $\Delta\to N\pi$ process is improved from PB(amd, jam) to PB(amd), while the $\Delta$ production shown in Fig.~\ref{fig:deltaprod} does not depend on this improvement in the decay of $\Delta$.
 To understand this behavior, we show in the right panel of Fig.~\ref{fig:pinum} the numbers of $\pi^-$ and $\pi^+$ in the final state.  We find that the number of $\pi^{+}$ is reduced more strongly than $\pi^{-}$ when the Pauli blocking is improved from PB(amd, jam) to PB(amd).  The following discussion will show that this is because $\pi^-$ production through the $\Delta^0\to p\pi^-$ process is not so strongly suppressed by the Pauli blocking, compared to the other $\Delta$ decay channels, as seen in Fig.~\ref{fig:deltadecay}.

%----------------------------
 
Here, in order to understand the Pauli blocking effect for pion production via $\Delta$ decay, 
we show the rates of $\Delta^- \rightarrow n \pi^-$ and $\Delta^{++} \rightarrow p \pi^{+}$ in the left panel of Fig.~\ref{fig:deltadecay} for different blocking options. 
The $\pi^-$ and $\pi^+$ are mainly produced by these reaction channels.
We show in this figure the integrated and normalized decay rates defined by
\begin{align}
\Delta^{++}\to p\pi^+ &:\quad  \frac{ \int_{0}^{\infty} R(\Delta^{++} \rightarrow p \pi^{+}) dt}{ \int_{0}^{\infty} N(\Delta^{++}) dt},
\label{Delta_pp}\\
\Delta^{-}\to n\pi^- &:\quad  \frac{ \int_{0}^{\infty} R(\Delta^{-} \rightarrow n \pi^{-}) dt}{ \int_{0}^{\infty} N(\Delta^-) dt}, 
\label{Delta_m}
\end{align}
where the numerator is the integrated number of the pion production by $\Delta\to N\pi$.  To make it easier to compare the effect among different Pauli blocking options, the number of the pion production is normalized by the time integral of the number of existing $\Delta$ resonances in Eqs.~(\ref{Delta_pp}) and (\ref{Delta_m}).
In addition, for the $\pi^{-}$ and $\pi^{+}$ production through the $\Delta^+$ and $\Delta^0$ decays, we show the following quantities in the right panel of Fig.~\ref{fig:deltadecay}, 
\begin{align}
\Delta^{+} \to n\pi^+ &:\quad  \frac{ \int_{0}^{\infty} R(\Delta^{+} \rightarrow n \pi^{+}) dt}{ \int_{0}^{\infty} N(\Delta^+) dt}, 
\label{Delta_p}\\
\Delta^{0}\to p\pi^- &:\quad  \frac{ \int_{0}^{\infty} R(\Delta^{0} \rightarrow p \pi^{-}) dt}{ \int_{0}^{\infty} N(\Delta^0) dt}.
\label{Delta_0}
\end{align}
A factor 3 has been multiplied in the right panel to compensate the decay branching factor $\tfrac13$ for these decay channels.  In both figures, it is generally observed that the pion production is strongly suppressed when the blocking in $\Delta\to N\pi$ is improved from PB(amd, jam) to PB(amd).
This is because Pauli blocking in PB(amd) is stronger than in PB(amd, jam).  We keep concentrating on the change from PB(amd, jam) to PB(amd), though we notice that the statements here and below also apply to the behaviors when the blocking is improved from PB($\tfrac14$jam) to PB(jam) in these figures.

In a closer view of each of the left and right panels of Fig.~\ref{fig:deltadecay}, according to the change from PB(amd, jam) to PB(amd),
the pion production is more strongly suppressed when the produced pion is accompanied by a neutron in the final state ($\Delta^- \rightarrow n \pi^-$ and 
$\Delta^{+} \rightarrow n \pi^+$)
 than in the case with a proton in the final state ($\Delta^{++} \rightarrow p \pi^+$ and $\Delta^0 \rightarrow p \pi^-$), which is because of the neutron-rich environment in the present system.
Since the decays of $\Delta^-$ and $\Delta^{++}$ are the dominant channels for the production of charged pions, one would expect that the stronger Pauli blocking would result in a stronger suppression of the $\pi^-$ production via $\Delta^-\to n\pi^-$, and therefore a lowering of $\pi^-/\pi^+$.
To the contrary, 
we find in Fig.~\ref{fig:pinum} that 
the number of $\pi^{+}$ is reduced more strongly than $\pi^{-}$ and the $\pi^-/\pi^+$ ratio increases, when the Pauli blocking is improved.
This suggests that the other channels of the $\Delta$ decay are playing some important roles.
In fact, we find in the right panel of Fig.~\ref{fig:deltadecay} that the suppression of the $\pi^-$ production via $\Delta^0 \rightarrow p \pi^-$ is relatively weak.  In particular, this suppression of $\Delta^0\to p\pi^-$ in the right panel is weaker than that of $\Delta^{++}\to p\pi^+$ in the left panel, even though the Pauli blocking is for a proton in both cases.  The same is qualitatively seen in the comparison of $\Delta^+\to n\pi^+$ in the right panel and $\Delta^-\to n\pi^-$ in the left panel for the decays with a neutron in the final state.  The origin of the difference between these two kinds of cases can be explained as follows.

The case for $\Delta^{++}$ (or $\Delta^-$) is relatively simple because the decay channel $\Delta^{++}\to p\pi^+$ competes only with the absorption channel $\Delta^{++}n\to pp$.  For an existing $\Delta^{++}$, the probability of its decay to produce a $\pi^+$
\begin{equation}
\frac{R(\Delta^{++}\to p\pi^+)}{R(\Delta^{++}n\to pp)+R(\Delta^{++}\to p\pi^+)}
\end{equation}
will be reduced by the improvement of the Pauli blocking from PB(amd, jam) to PB(amd), simply because the decay rate $R(\Delta^{++}\to p\pi^+)$ is suppressed.  On the other hand, for a $\Delta^0$ (or $\Delta^+$) resonance, the decay channel of $\Delta^0\to p\pi^-$ competes with more channels, so the chance of this decay is
\begin{equation}
\frac{R(\Delta^0\to p\pi^-)}{R(\Delta^0N\to NN)+R(\Delta^0\to n\pi^0)+R(\Delta^0\to p\pi^-)}.
\end{equation}
This is of course reduced when the Pauli blocking for protons suppresses $R(\Delta^0\to p\pi^-)$, but also it tends to be increased when $R(\Delta^0\to n\pi^0)$ in the denominator is suppressed.  Namely, if the channel of $\Delta^0\to n\pi^0$ is closed by the strong blocking of neutrons, the probability flows to other channels including the decay channel of $\Delta^0\to p\pi^-$.  This is why the $\Delta^0\to p\pi^-$ decay in Fig.~\ref{fig:deltadecay}(right) is not suppressed so strongly by the change from PB(amd, jam) to PB(jam), compared to the other $\Delta\to N\pi$ decay channels.  This eventually results in the increase of the $\pi^-/\pi^+$ ratio in Fig.~\ref{fig:pinum} when the blocking is changed from PB(amd, jam) to PB(amd).
% 

%------------------------------------------------

\subsection{Discussions on PB(amd-h)}

Using the Husimi function as the Pauli blocking probability, as in the blocking option PB(amd-h), may be an attractive idea because the Husimi function $f=f_{\text{amd-h}}^\tau$ defined by Eq.~\eqref{eq:amd-husimi} has a good property as a probability, $0\le f\le 1$.  However, it has been smeared and therefore it does not agree with the true distribution of particles.  As already seen in the middle and right panels of Fig.~\ref{fig:f_NNND}, the blocking probability in PB(amd-h) is lower than that in PB(amd) in the phase-space region that is important for the Pauli blocking e.g.\ in the $NN\to N\Delta$ process.  In Fig.~\ref{fig:deltaprod} and other figures, the results in PB(amd-h) are shown by the rightmost isolated points.  All results are consistent with the idea that the Pauli blocking in PB(amd-h) is somewhat weaker than in PB(amd).  On the other hand, the Husimi function $f_{\text{amd-h}}^\tau$ is supposed to be equivalent to the quantity $f_{\text{jam}}^\tau$ in the JAM code [Eq.~\eqref{eq:f-jam}] in the sense that both are smeared distributions.  In fact, the left and the right panels of Fig.~\ref{fig:f_NNND} showed that $\langle f_{\text{amd-h}}^\tau\rangle$ is almost identical with $\langle f_{\text{jam}}^\tau\rangle$.  However, the Husimi function has an advantage that it is free from unphysical fluctuations due to test particle sampling, and therefore it can be used as the blocking probability without truncation.  Consequently, we have a relation $\langle f_{\text{amd-h}}^\tau\rangle > \langle\min(f_{\text{jam}}^\tau,1)\rangle$ for the actual blocking probability.  The results shown in Fig.~\ref{fig:deltaprod} and others are, in fact, consistent with the weaker blocking in PB(jam) compared to PB(amd-h).

\section{Summary\label{sec:summary}}
 In this paper, we have investigated important effects of Pauli blocking in the productions of $\Delta$ resonances and pions in heavy-ion collisions of neutron-rich nuclei (${}^{132}\mathrm{Sn}+{}^{124}\mathrm{Sn}$) at 270 MeV/nucleon, using different Pauli blocking (PB) options in the AMD+JAM approach.  The charged pion ratio $\pi^-/\pi^+$ is considered to be one of the important quantities to constrain the symmetry energy, and thus precise predictions are required.  The present work aimed to minimize the inaccuracy in the Pauli blocking, particularly for the $NN\to N\Delta$ and $\Delta\to N\pi$ processes, by improving the method in the model.  The most standard method of Pauli blocking, which we call PB(jam), suffers from the problem of insufficient blocking due to unphysical fluctuations and additional smearing, as observed in the transport code comparison of Ref.~\cite{yxzhang2018} for an almost degenerate Fermi gas. This general and fundamental problem in QMD codes can be overcome in the AMD+JAM approach, in PB(amd) option, by faithfully using the Wigner function calculated from the AMD wave function in the AMD code, which is the most reliable treatment in our approach.

We found that the blocking in $NN\to N\Delta$ and $\Delta\to N\pi$ processes can never be ignored in predicting the $\Delta$ and pion quantities, and blocking methods faithful to the AMD Winger function significantly
change the results.  With the more accurate and therefore stronger Pauli blocking, the $\Delta^-/\Delta^{++}$ and $\pi^-/\pi^+$ ratios become higher, for the productions of these particles.  The effects of blocking for these productions are mostly understood based on the strong blocking for neutrons, compared to protons, in the neutron-rich environment.  Especially, it is straightforward to understand the effect in the $\Delta$ production ($NN \rightarrow N \Delta$) process.  On the other hand, the effect in the $\Delta\to N\pi$ process is counter-intuitive, but we have understood it by considering the competition of all channels for $\Delta$.  We also found that the amount of the change in the $\pi^-/\pi^+$ ratio from PB(jam) to PB(amd) is comparable to the sensitivity to the symmetry energy.

At present, we have a good Pauli blocking treatment which uses the Wigner function for the AMD wave function. However, the Wigner function $f$ can sometimes become $f<0$ or $f>1$, and therefore is not always suitable as the blocking probability.  If one find a more precise approach which do not require unnatural truncation of the probability, predictions e.g.\ of the $\pi^-/\pi^+$ ratio may be further improved.

\begin{table}
\caption{\label{tab:auau} Charged pion muliplicities and the $\pi^-/\pi^+$ ratio in Au + Au central collisions at 400 MeV/nucleon.  The results of AMD+JAM calculations, with different Pauli blocking options PB($\tfrac14$jam), PB(jam) and PB(amd), are compared with the FOPI experimental data \cite{reisdorf2010}.  The AMD calculation was made with a soft symmetry energy ($L=46$ MeV) and with cluster correlations.}
\begin{tabular}{ccccc}
\hline\hline
& PB($\tfrac14$jam) & PB(jam) & PB(amd) & FOPI data \\
\hline
$\pi^-$ & $2.228\pm0.015$     & $2.012\pm0.011$ & $1.924\pm0.013$ & $2.80\pm0.14$\cite{reisdorf2010} \\
$\pi^+$ & $0.900\pm0.007$      & $0.764\pm0.004$ & $0.701\pm0.004$ & $0.95\pm0.08$\cite{reisdorf2010} \\
$\pi^-/\pi^+$ & $2.48\pm0.02$ & $2.64\pm0.02$ & $2.74\pm0.02$ & $2.95\pm0.29$\footnote{An experimental uncertainty $\pm0.29$ for $\pi^-/\pi^+$ is estimated by propagating the errors in the multiplicities of $\pi^-$ and $\pi^+$, assuming that they are independent.} \\
\hline\hline
\end{tabular}
\end{table}

The aim of the present paper was to theoretically study and improve Pauli blocking, without direct comparision with experimental data.  Comparisons with predictions by other transport models and with the experimental data by the S$\pi$RIT collaboration are currently in progress \cite{jhang2020} for Sn + Sn systems at 270 MeV/nucleon.  In Au + Au collisions at 400 MeV/nucleon for which the FOPI experimental data are available \cite{reisdorf2007,reisdorf2010}, the result of AMD+JAM changes as shown in Table \ref{tab:auau} when the Pauli blocking treatment is improved from PB($\tfrac14$jam) to PB(jam) and then to PB(amd), in a way similar to the Sn + Sn system studied in the present paper.  With the best blocking method PB(amd), the calculated result of the $\pi^-/\pi^+$ ratio with a soft symmetry energy and with cluster correlations may agree with the FOPI data within experimental uncertainties.  However, the absolute value of pion multiplicity has not been carefully studied yet, e.g.\ by considering in-medium effects in the $\Delta$ production and absorbtion cross sections.  
In addition, we should keep in mind that other model ingredients, such as the threshold for the $\Delta$ resonance production in nuclear medium \cite{ferini2005,xu2010,xu2013,song2015,zhenzhang2017,zhenzhang2018,cozma2016,cozma2017} and the pion optical potential, also affect pion observables.

\section*{Acknowledgments}
The computation was carried out at the HOKUSAI supercomputer system of RIKEN.
This work was supported by JSPS Overseas Research Fellowships and JSPS KAKENHI Grant Numbers JP19K14709, JP17K05432, JP19H01898 and JP19H05151.

\bibliography{ono_nucl}

\end{document}